\documentclass[preprint2]{aastex6}
\usepackage{amsmath}
\usepackage{float}

\slugcomment{Accepted by The Astrophysical Journal Letters on June 12, 2017}

\begin{document}

\title{Accretion-Induced Collapse From Helium Star + White Dwarf Binaries}

\author{Jared Brooks\altaffilmark{1}, Josiah Schwab\altaffilmark{2,3,4}, Lars Bildsten\altaffilmark{1,5}, Eliot Quataert\altaffilmark{2}, Bill Paxton\altaffilmark{5}}

\altaffiltext{1}{Department of Physics, University of California, Santa Barbara, CA 93106}

\altaffiltext{2}{Astronomy and Physics Departments and Theoretical Astrophysics Center, University of California, Berkeley, CA 94720, USA}
\altaffiltext{3}{Department of Astronomy and Astrophysics, University of California, Santa Cruz, CA 95064, USA}
\altaffiltext{4}{Hubble Fellow}
\altaffiltext{5}{Kavli Institute for Theoretical Physics, Santa Barbara, CA 93106}

\begin{abstract}

Accretion-induced collapse (AIC) occurs when an O/Ne white dwarf (WD) grows to nearly the Chandrasekhar mass ($M_{\rm Ch}$), reaching central densities that trigger electron captures in the core.  
Using Modules for Experiments in Stellar Astrophysics (\texttt{MESA}), we present the first true binary simulations of He star + O/Ne WD binaries, focusing on a $1.5 M_\odot$ He star in a 3 hour orbital period with $1.1-1.3 M_\odot$ O/Ne WDs.  
The helium star fills its Roche lobe after core helium burning is completed and donates helium on its thermal timescale to the WD, $\dot{M}\approx3\times10^{-6} M_\odot$/yr, a rate high enough that the accreting helium burns stably on the WD.
The accumulated carbon/oxygen ashes from the helium burning undergo an unstable shell flash that initiates an inwardly moving carbon burning flame. 
This flame is only quenched when it runs out of carbon at the surface of the original O/Ne core.  
Subsequent accumulation of fresh carbon/oxygen layers also undergo thermal instabilities, but no mass loss is triggered, allowing $M_{\rm WD}\rightarrow M_{\rm Ch}$, triggering the onset of AIC.
We also discuss the scenario of accreting C/O WDs that experience shell carbon ignitions to become O/Ne WDs, and then, under continuing mass transfer, lead to AIC. 
Studies of the AIC event rate using binary population synthesis should include all of these channels, especially this latter channel, which has been previously neglected but might dominate the rate.

\end{abstract}


\section{Introduction}

White dwarfs (WDs) that are primarily composed of oxygen and neon (O/Ne) in their cores are expected to collapse to form neutron stars (NSs) as they approach the Chandrasekhar mass ($M_{\rm Ch}$). 
The collapse is triggered by the onset of electron capture reactions in the center of the star that occur above a critical density; this is the same trigger that gives rise to electron capture supernovae in single star evolution \citep{Nomoto1979, Miyaji1980, Nomoto1984, Nomoto1987, Takahashi2013}, with the key difference that the AIC scenario lacks an extended stellar envelope.
When the WD grows in mass via accretion from a binary companion, this process is referred to as accretion-induced collapse \citep[AIC;][]{Canal1990, Nomoto1991, Woosley1992, Ritossa1996, Dessart2006, Metzger2009, Darbha2010, Piro2013, Tauris2013}.
AIC supernovae are predicted to be very fast and faint \citep{Woosley1992, Dessart2006} and thus difficult to observe. 
The remnant NSs may, however, be detectable as low-mass binary pulsars \citep{Nomoto1991}, or, if they are later spun up by accretion, millisecond pulsars (MSPs) \citep{Tauris2013}.

In this paper, we study AIC progenitor scenarios that involve a He burning star as the donor.
The mass transfer rates of $\dot{M}\approx3\times10^{-6} M_\odot/$yr experienced by these systems overlap with the steady helium burning rates for WDs, allowing for steady growth of the WD cores up to $M_{\rm Ch}$ \citep{Yoon2003}.
We find that the C/O ashes from the steady helium burning shell ignite unstably in a shell flash, but that these carbon burning episodes do not significantly interrupt the growth of the WD, so that all models grow to $M_{\rm Ch}$.
This work presents the first modeling of the carbon shell flashes from steady helium burning in relation to growing the cores of O/Ne WDs.

In this work, we model the binary evolution of a He star + WD system by simultaneously modeling the evolution of the binary system, including the structure of both stars and their orbit.
In addition to this more detailed model of the standard He star + O/Ne WD channel, we also speculate that He-star+C/O models that undergo carbon shell ignition before carbon core ignition can be a channel for AIC, as discussed in \cite{Brooks2016}.
A core ignition of a C/O WD would lead to a SN Ia, but a shell ignition non-explosively transforms the C/O WD into an O/Ne WD, which continues to accrete until AIC is achieved.

In \S \ref{sec:grow} we discuss the initial parameter and modeling assumptions of our binary evolution $\texttt{MESA}$ calculations.
In \S \ref{sec:ONe-accretors} we follow the growth of an O/Ne WD through stable helium shell burning, and explain the physics of the unstable carbon shell burning episodes that occur.
We explore the C/O WD carbon shell ignition to O/Ne WD AIC channel in \S \ref{sec:CO-accretors}, and discuss the structure of the WD leading up to the AIC event in \S \ref{sec:structure}.
We conclude in \S \ref{sec:concl} by highlighting the likely impact on expected rates.

\section{Binary Evolution and Mass Transfer}\label{sec:grow}

We use \texttt{MESA} (r7624) \citep{Paxton2011, Paxton2013, Paxton2015a} to model the full set of stellar structure equations for both stars simultaneously; we also model the evolution of the binary parameters taking into account their interaction through mass transfer.
Both the He star and WD are created in $\texttt{MESA}$ in single star evolution by starting with a zero-age main sequence (ZAMS) star.
We generated models of O/Ne WDs using the same set of physical assumptions as \cite{Farmer2015} from initial masses of $11-12 M_\odot$ and removed the envelopes once carbon burning had ended.
The He star starts as an $8.5 M_\odot$ star and is evolved until just before He core ignition, at which point the H envelope is artificially removed. 
Both models had solar metallicity. 
After we remove the envelope from the WD, we let it cool for 10 Myr (roughly the difference in main-sequence lifetimes), then place it in a binary calculation with the He star.

\begin{figure}[H]
  \centering
  \includegraphics[width = \columnwidth]{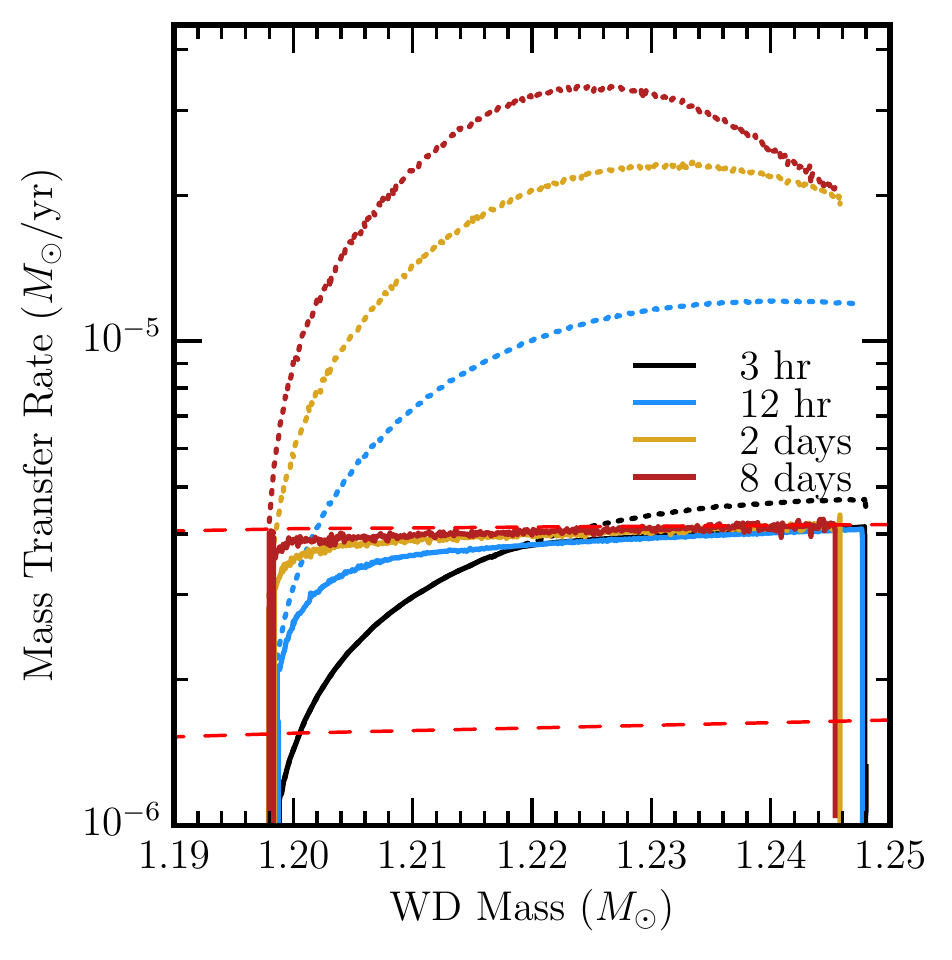}
  \caption{\footnotesize The mass transfer rates of $1.5 M_\odot$ He star + WD systems with a range of initial orbital periods from 3 hours to 8 days.
  The dotted lines show the rate at which mass is removed from the He star ($\dot{M}_{\rm He}$); the solid lines show the rate at which mass is accepted by the WD ($\dot{M}_{\rm WD}$).
  The difference is assumed to be lost from the system ($\dot{M}_{\rm wind}$), carrying the specific angular momentum of the WD accretor.
  The stable helium burning boundaries are shown by the dashed red lines.
  These simulations are only followed up to the first carbon shell ignition.}
  \label{fig:15}
\end{figure}

We model the growth of the mass of the WD cores using the same method as growing the core mass of C/O WDs in \citet{Brooks2016}, which calculates the fraction of the donated mass the accretor can retain based on its maximum rate of He-burning. 
While the system is in the regime for steady He-burning, the WD burns helium to carbon and oxygen at the same rate that it is accreting helium. 
As the mass transfer rates rise above the steady burning regime, the WD rapidly expands into its Roche lobe.
We assume the the WD only accepts mass at the maximum steady burning rate and that the rest of the mass is lost from the system such that $\dot{M}_{\rm WD} + \dot{M}_{\rm wind} = \dot{M}_{\rm He}$, where $\dot{M}_{\rm wind}$ is the rate of mass loss from the binary, $\dot{M}_{\rm He}$ is the mass loss rate of the helium donor star, and $\dot{M}_{\rm WD}$ is the mass gain rate of the WD. 
We assume that the wind carries with it the specific angular momentum of the WD accretor.
Figure \ref{fig:15} shows  the resulting evolution for a range of initial orbital periods.
This shows that as the initial orbital period increases, the mass transfer rates increase, and, thus, the mass loss and angular momentum loss rates from the system increase.
If our prescription underestimates the specific angular momentum removed by the wind, the longer initial orbital period systems may be subject to mergers.
Therefore, we choose 3 hours as our fiducial orbital period, as these short period systems have the lowest wind mass loss rates, and are the most likely to avoid mergers and be accurately modeled (given our assumptions) in the event that we underestimated the specific angular momentum of the wind.
As the accretion rates (solid curves) do not vary much between the given range of orbital periods after gaining the first $0.02 M_\odot$, the structure of the WDs should be qualitatively similar as $M_{\rm WD}\rightarrow M_{\rm Ch}$, relatively independent of the initial orbital period \footnote{The algorithm we use to determine the fraction of donated mass that stays on the WD leads to a small discrepancy for the 12 hour orbital period system, see \cite{Brooks2016} a discussion.}.

The high mass transfer rates and large core masses considered here lead to the ignition of carbon in the shell of helium burning ashes.
Due to heat from the ashes leaking into the colder core, the first unstable ignition in this shell occurs off-base.  
This leads to the formation of an inwardly-propagating carbon flame.  
In the case of an O/Ne WD accretor (see \S~\ref{sec:ONe-accretors}), this carbon flame will propagate inwards until it runs out of carbon to burn at the edge of the O/Ne core.  
In the case of a C/O WD accretor (see \S~\ref{sec:CO-accretors}), since the whole WD is C/O, we expect that the carbon burning flame will propagate all the way to the center over about a year, converting the C/O WD to a O/Ne WD.

\begin{figure}[H]
  \centering
  \includegraphics[width = \columnwidth]{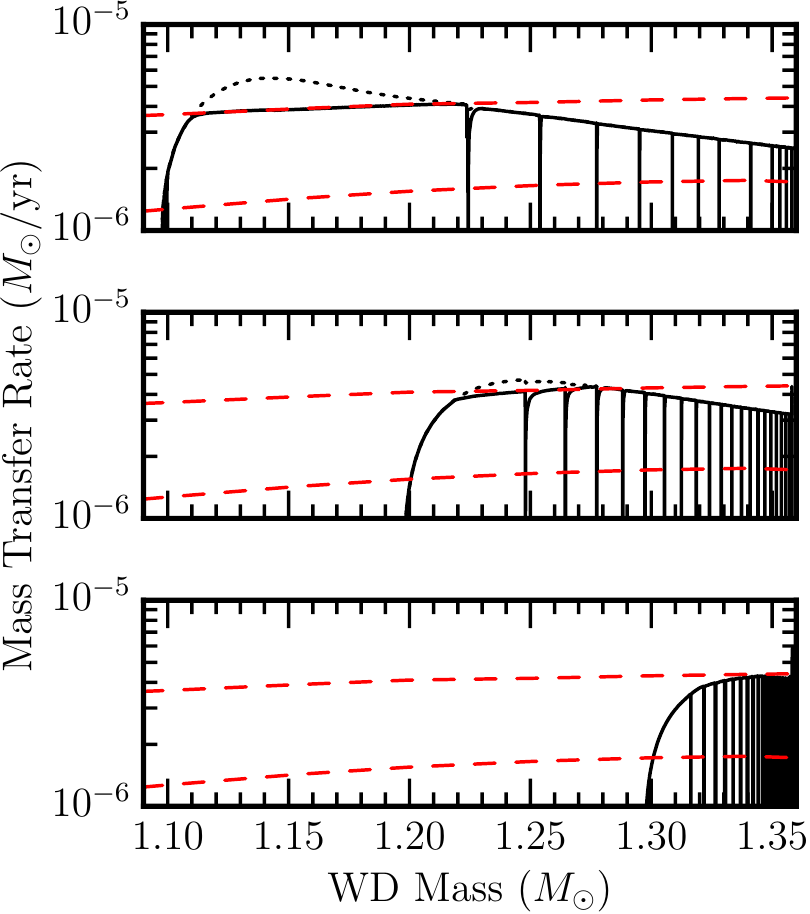}
  \caption{\footnotesize Mass transfer rates for O/Ne WDs with initial masses of  $1.1, 1.2,$ and $1.3 M_\odot$ in a binary system with a $1.5 M_\odot$ He star (solid black) at an initial orbital period of 3 hours.
  The mass transfer is punctuated by brief mass loss eposides caused by carbon shell flashes in the helium burning ashes.
  The solid tracks are the rate at which the WD is gaining mass; the dotted tracks are the rate at which the He star is losing mass. 
  The difference between the dotted and solid tracks represent the mass that is lost from the system. 
  The stable helium burning boundaries are shown by the dashed red lines.}
  \label{fig:8}
\end{figure}

In both cases, since we still have a massive WD core and high accretion rates, the WD will continue to build up C/O ashes from steady helium burning, and burn the carbon in short shell flash episodes.
All subsequent C/O layers ignite at the base of the freshly accumulated carbon. 
This is analogous to thermally pulsing AGB stars, where the hydrogen burning layer supplies helium ashes at below the steady helium burning rate, so the underlying helium burning shell is thermally unstable and will thus pulse.
An important difference is that for these carbon burning episodes, most of their energy output is balanced by thermal neutrinos emitted in the convective regions above the burning layer \citep{Timmes1994}, meaning that negligible mass is lost (${\lesssim}10^{-5}M_\odot$) per carbon shell flash. 
This leads to a punctuated mass transfer history, as shown in Figure \ref{fig:8}, where steady mass accretion is repeatedly interrupted during the brief carbon flashes.
The duty cycle of these carbon flashes is about 0.001.

As we are only interested in showing the binary conditions needed to reach AIC, we halt evolution of all models when $\log\rho > 9.6$.
At this point electron capture reactions in the center of the star begin to significantly remove pressure support \citep{Schwab2015a}, and the timescale to collapse becomes hundreds of years.

\section{O/Ne WD Accretors}
\label{sec:ONe-accretors}

We now discuss the details of the evolution of an initially $1.2 M_\odot$ O/Ne WD as it experiences its first carbon shell flash, an ingoing carbon flame, and then subsequent carbon flashes.  
The model starts with $0.025 M_\odot$ of C/O above a $1.175 M_\odot$ O/Ne core, which results from helium shell burning during the progenitor's AGB phase \citep{Gil?Pons2001}.

\subsection{First carbon shell flash}

\begin{figure}[H]
  \centering
  \includegraphics[width = \columnwidth]{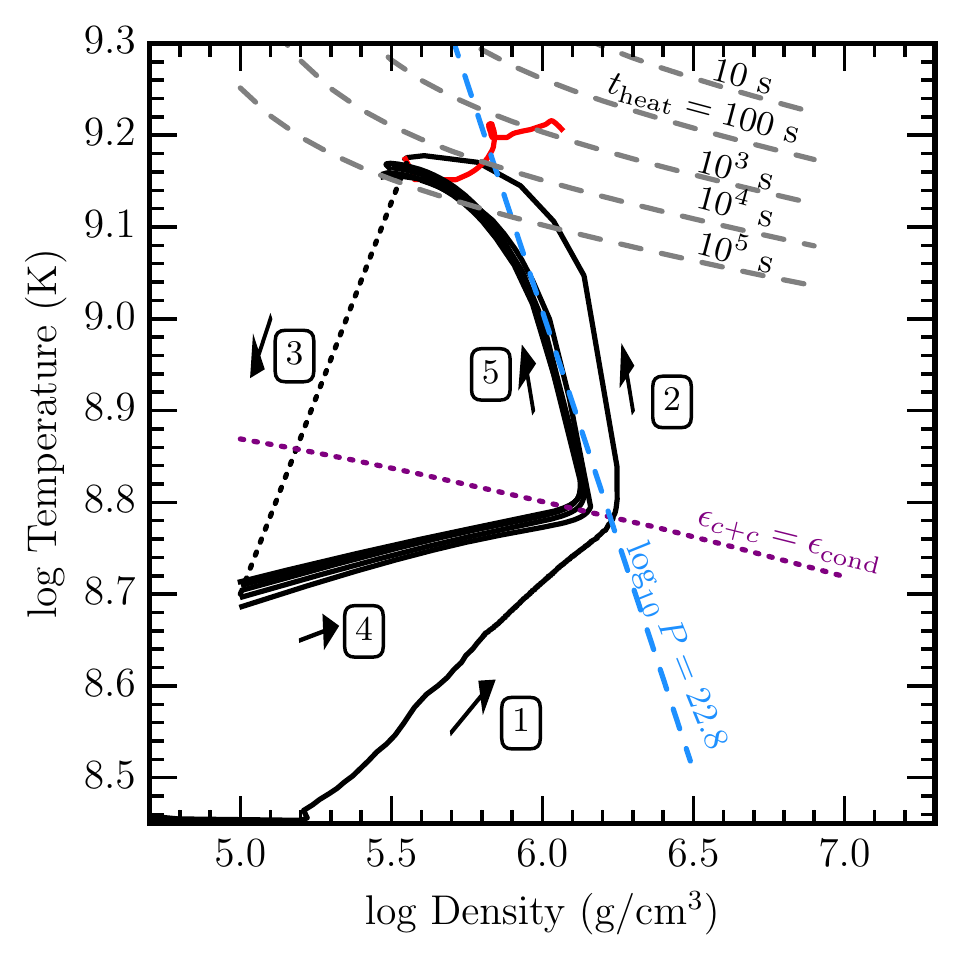}
  \caption{\footnotesize The evolution of the density and temperature at the mass coordinate of maximum temperature in the carbon layer. 
  As accretion starts, this begins in the bottom-left corner and evolves along arrow 1. 
  After crossing the purple dotted $\epsilon_{\rm C+C}=\epsilon_{\rm cond}$ line, carbon burning is ignited and it evolves along arrow 2. 
  The evolution during the carbon flame phase (\S \ref{sec:theflame}) is colored red.
  When carbon burning is quenched and the carbon layer is no longer well defined, we schematically represent the evolution with arrow 3.
  We begin to visualize the new carbon layer when $\log\rho=5$. 
  Then, as helium burning deposits more mass in the C/O layer, evolution proceeds along arrow 4 until carbon is ignited, at which point the evolution proceeds along arrow 5 until carbon burning quenches, and then the cycle repeats.
  The upper dashed lines represent locations where the heating time is a fixed value.}
  \label{fig:t-vs-rho}
\end{figure}

Figure \ref{fig:t-vs-rho} shows the evolution of the density and   

\begin{figure}[H]
  \centering
  \includegraphics[width = \columnwidth]{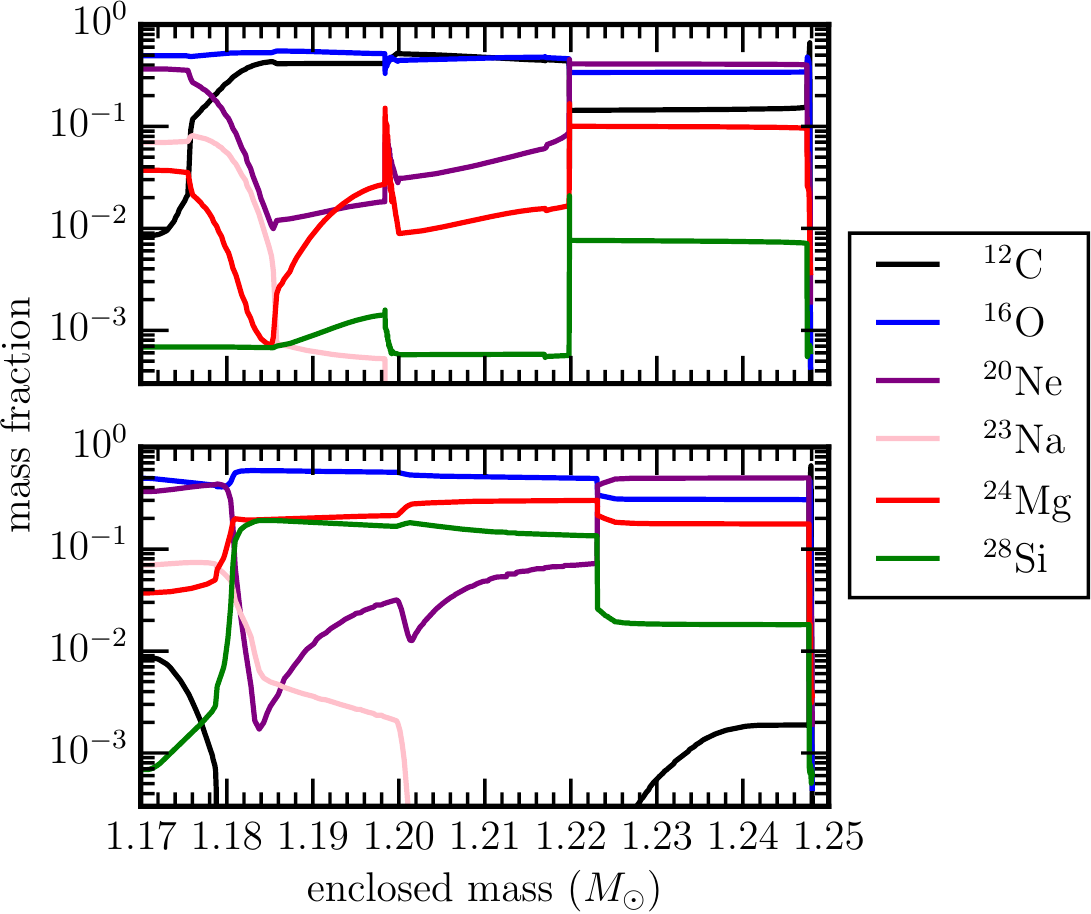}
  \caption{\footnotesize Top panel: Composition profile at minimum $t_{\rm heat}$ of the first carbon shell flash. 
  The starting model has a mass of $1.20 M_\odot$, with $0.025 M_\odot$ of C/O on top of a $1.175 M_\odot$ O/Ne core. 
  Additional C/O is deposited from steady helium burning ashes until the mass of the WD reaches $1.247 M_\odot$. 
  The ignition of carbon burning starts at the mass coordinate $M_r=1.220 M_\odot$.
  Bottom panel: Composition profile after the carbon burning flame is quenched. 
  }
  \label{fig:12}
\end{figure}

\noindent temperature at the mass coordinate of the maximum temperature in the carbon layer.  
This begins in the bottom-left corner and evolves (along arrow 1) towards higher temperatures and densities as accretion and steady helium burning increase the mass of the C/O layer.   
The carbon will unstably ignite when the C+C energy generation rate exceeds the rate at which conduction can remove the heat from a region with a local thickness of a scale height($\epsilon_{\rm C+C}>\epsilon_{\rm cond}$, \citealt{Cumming2001}; above the purple dotted line).  
The first ignition starts in the middle (as opposed to the base) of the upper carbon layer (the ashes), as the compressional heat and leftover heat from helium burning in the C/O layer leaks through to the colder material below.  
The first shell flash starts at a mass coordinate of $M_r=1.220 M_\odot$ with a $0.045 M_\odot$ layer of C/O below it, as shown in Figure \ref{fig:12}.

Once the carbon burning at the ignition location is generating heat faster than conduction can remove the heat, a thermonuclear runaway results.
Since the ignition of carbon burning occurs in a thin shell, the thermonuclear runaway proceeds initially at roughly constant pressure (along arrow 2 in Fig. \ref{fig:t-vs-rho}).
The deviation from constant pressure occurs when the shell becomes radially extended so that continued entropy production mostly leads to density reduction.
This keeps the runaway mostly hydrostatic, as evidenced by the fact that the heating time ($t_{\rm heat} = c_pT/\epsilon_{C+C}$) remains substantially below the local dynamic time, $t_{\rm dyn}=H/C_s$, where $H$ is the local pressure scale height and $C_s$ is the sound speed; $t_{\rm dyn}$ never gets much greater that $0.1$ seconds.
About a day  after the ignition, the carbon above the ignition location is depleted to a mass fraction ${\approx}0.001$.

\subsection{The carbon flame}
\label{sec:theflame}

The layer directly beneath the burning layer heats up, primarily via electron conduction from the burning layer above, until it is hot enough to sustain its own carbon burning. 
In turn, it heats the underlying layer, and a slow, inwardly-propagating deflagration wave results.
The temperature and density of the burning front are shown in Figure \ref{fig:t-vs-rho} in the red part of the curve that starts at $\log\rho=5.6$, $\log T=9.2$ and ends at $\log\rho=6.45$, $\log T=9.2$.
The minimum heating timescale reaches $t_{\rm heat}\approx100$ seconds during the flame duration.
Therefore, the burning is always hydrostatic.
The flame is thin (on the order of a kilometer or less) with a steep temperature gradient due to the high temperature sensitivity of carbon burning.

Ahead of the flame front, about two-thirds of the energy absorbed by unburned material goes into heating, and the other one-third is used for expanding. 
In the convective region behind the flame, energy release from carbon burning is mostly balanced by neutrino emission \citep{Timmes1994}.
Some energy in this region raises the entropy of the material, 
leaving very little energy available to emerge from the C/O core \citep{Garcia-Berro1997}.

\begin{figure}[H]
  \centering
  \includegraphics[width = \columnwidth]{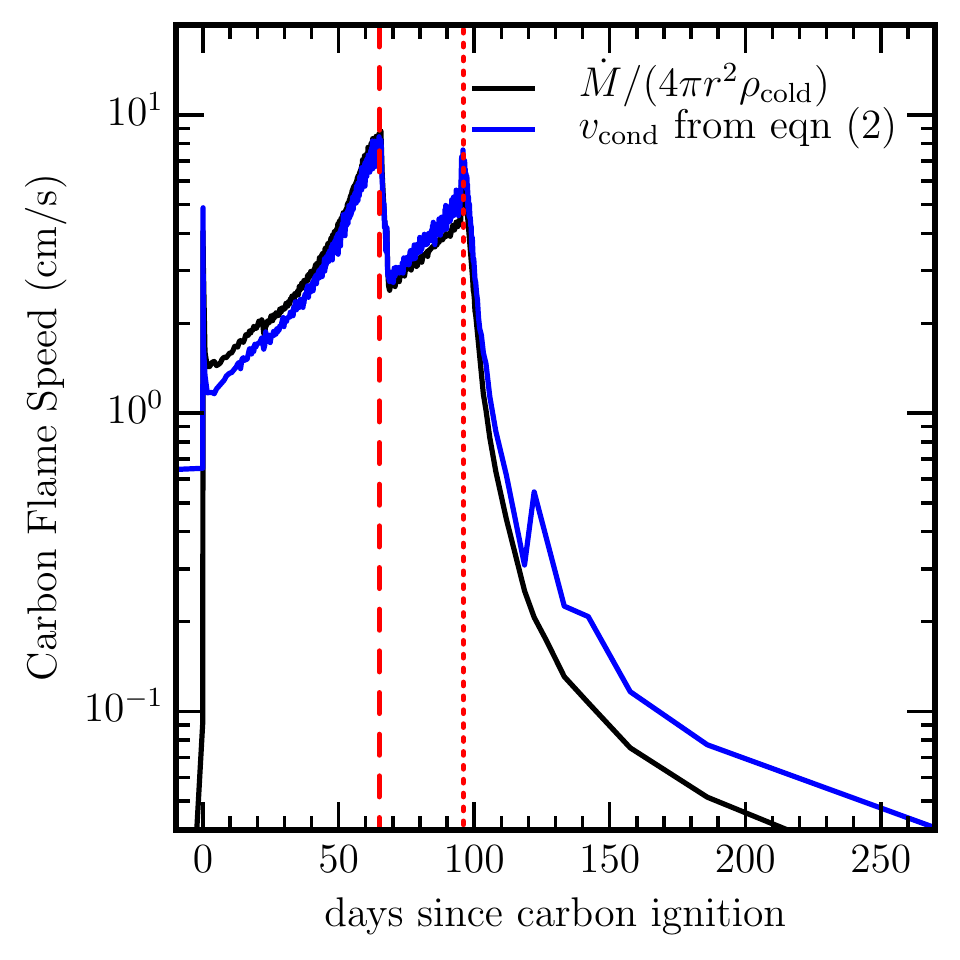}
  \caption{\footnotesize The numerical Lagrangian flame speed for the $M_{\rm WD}=1.2 M_\odot$ case is shown by the black solid line, and the theoretical flame speed, using equation \ref{eqn:5} and the relevant local quantities, including the fitted prefactor of $0.25$, is shown by the solid blue line.
  The vertical dashed red line labels the time when the flame has propagated all the way through the C/O ashes deposited from steady helium burning and enters into the C/O layer that existed on top of the O/Ne left over from formation. 
  The flame speed slows slightly because this material is colder and denser than the C/O above it.
  The vertical dotted red line labels the time when the flame has propagated through the WDs original C/O layer and begins to quench as it runs out of carbon to burn within the O/Ne WD.
}
  \label{fig:10}
\end{figure}

We can roughly estimate a speed by $v\sim l/t_{\rm heat}$.
We have the heating timescale, $t_{\rm heat}$, in the burning front, and define a geometrical thickness of the shell, $l$, by,
\begin{equation}\label{eqn:4} \epsilon_{\rm C+C} \approx \dfrac{D}{\rho}\dfrac{\partial^2 T}{\partial r^2} \approx \dfrac{D}{\rho}\dfrac{T_b}{l^2} , \end{equation}
where $D$ is the thermal diffusion coefficient and $T_b$ is the temperature at the base of the burning layer.
Plugging in these values at the time when the flame is at a mass coordinate of about $1.21 M_\odot$ yields a flame speed within an order of magnitude of the numerical flame speed given by the model.
Following the more rigorous approach of \citet{Timmes1994} gives,
\begin{equation}\label{eqn:5} v \approx 0.25\left(\dfrac{c\epsilon_{\rm C+C}}{\kappa \rho_{\rm cold} E}\right)^{1/2} , \end{equation}
where $0.25$ is derived from a rough fit to the numerical results, $c$ is the speed of light, $\kappa$ is the opacity at the flame front, $\rho_{\rm cold}$ is the density on the cold (inner) side of the flame front, and $E (=c_pT)$ is the internal thermal energy per gram. 
Using the values recorded at the location of maximum $\epsilon_{\rm C+C}$, we plot the flame speeds predicted by equation \ref{eqn:5}, along with the numerical flame speed in the model given by $v=\dot{M}/(4\pi r^2\rho)$ for the $M_{\rm WD}=1.2 M_\odot$ case in Figure \ref{fig:10}.
We use the Lagrangian speed $\dot{M}/(4\pi r^2\rho)$ because the Eulerian speed, $dR_{\rm max}/dt$, where $R_{\rm max}$ is the radius of the maximum burning location, is impacted by the underlying core expansion.
This Lagrangian speed is shown in solid black, and the theoretical flame speed from \cite{Timmes1994} is shown in blue. 
The flame speed steadily increases from a few cm s$^{-1}$ to a few dozen cm s$^{-1}$ in the first 65 days due to an increase in the temperature and carbon mass fraction at the burning front.
After 65 days, the flame reaches $M_r=1.20 M_\odot$ and the carbon mass fraction drops from $0.52$ to $0.41$, which reduces the energy generation rate, $\epsilon_{\rm C+C}$, by a factor of ${\approx}1/4$, which in turn lowers the flame speed by a factor of ($\Delta v_{\rm flame}\sim(\Delta\epsilon_{\rm C+C})^{1/2}$)$\approx0.4$. 
The shapes of the two curves match, meaning that the theoretical and numerical flame speed differ by a constant factor of less than order unity, which we set to 0.25.

The flame takes about 100 days to reach the mass coordinate $M_r=1.175 M_\odot$, where the carbon mass fraction drops from 0.4 to ${<}0.01$. 
The carbon flame only fully quenches after 200 years, which is not shown in Figure \ref{fig:10}. 
The composition profile after the carbon flame has been quenched is shown in Figure \ref{fig:12}. 
We can discern three separate regions processed by carbon burning. 
The outermost region, starting at $M_r=1.223 M_\odot$, was burned in one week and is dominated by $^{20}$Ne ($X_{\rm ^{20}Ne}=0.50$, $X_{\rm ^{16}O}=0.30$, $X_{\rm ^{24}Mg}=0.08$). 
This is due to the fact that the carbon in this region burns hot ($T_b\approx1.6\times10^9$ K) and fast (only about a day).
The shell processed by the (relatively) fast carbon flame ($1.20 M_\odot < M_r < 1.223 M_\odot$) is dominated by $^{16}$O ($X_{\rm ^{16}O}=0.55$, $X_{\rm ^{24}Mg}=0.30$, $X_{\rm ^{28}Si}=0.15$, $X_{\rm ^{20}Ne}=0.05$). 
This region burns at a comparable temperature ($1.6\times10^9<T_b<2.0\times10^9$ K), but the burning happens over a longer timescale, enough for the reactions $^{20}$Ne($\gamma, \alpha$)$^{16}$O and $^{20}$Ne($\alpha, \gamma$)$^{24}$Mg($\alpha, \gamma$)$^{28}$Si to deplete ${\sim}3/4$ of the $^{20}$Ne produced in carbon burning.
The inner shell ($1.175 M_\odot < M_r < 1.20 M_\odot$) is processed by the (relatively) slow carbon flame and is dominated by $^{16}$O (similar composition to layer above).

\subsection{Subsequent Carbon Shell Flashes}
\label{sec:theflashes}

After the carbon burning flame is quenched, the energy that was absorbed begins radiating outwards. 
As it reaches and expands the helium shell, the helium burning is shut off. 
During this transition, the base of the carbon layer and the maximum carbon burning location are not well defined, as all the existing carbon is consumed, so in Figure \ref{fig:t-vs-rho} the evolution is schematically represented by the dotted line in the direction of arrow 3.
The model then resumes helium accretion, which increases the mass of the helium layer until steady helium burning reignites, which builds a new C/O layer. 
We catch up with the evolution of the new C/O layer once its density reaches $10^5$ g cm$^{-3}$ (arrow 4). 
The O/Ne layer beneath the C/O layer is still hot, so that compressional heat generated in the C/O layer cannot effectively radiate inwards. 
This means that the C/O layer will ignite at its base (arrow 5), preventing any more inwardly-propagating carbon flames. 
The resulting pulsing behavior is analogous to the carbon flashes that occur on the surface of accreting neutron stars \citep{Cumming2001}.
Each subsequent carbon shell flash ignites on a more massive core, leading to lower ignition masses and somewhat lower minimum $t_{\rm heat}$. 
The changes in radius and luminosity during these shell flashes are of order unity, meaning they are difficult to observe and do not lead to appreciable mass loss.

\section{C/O WD accretors as AIC Progenitors}
\label{sec:CO-accretors}

In order to estimate the full contribution of the AIC channel with He star donors, one must include not only WDs that begin the accretion stage as O/Ne WDs, but also those C/O WDs that transform into O/Ne WDs \textit{during} the accretion stage. 
This can occur, as described in \cite{Brooks2016}, if the donor is massive enough to sustain high accretion rates such that the hot carbon ashes near the surface of the accretor ignite before a carbon core ignition occurs. 
As discussed in \S \ref{sec:theflame}, this shell ignition of carbon is non-explosive and initiates an inwardly propagating carbon flame much like that found in WD merger scenarios \citep{Nomoto1985, Saio1985, Saio1998, Schwab2016}.

An important contrast from the merger scenarios is that the C/O WDs in our scenario are more massive ($M>1.2M_\odot$)
This means that they have much larger core densities and the flames have higher bounding temperatures. 
This leads to thin flames of initial width $\sim 10^2$ cm, agreeing with \citet{Timmes1994} for the same conditions ($\rho\approx10^7$ g/cm$^3$, $T\approx1.5\times10^9$ K).
Using the flame speed at the start of the flame, we expect the entire core will be converted to O/Ne on a timescale of years.  However, propagating such an initially thin flame all the way to the center of a massive C/O WD
requires $\ga 100$x more timesteps than the other flames studied in this work.

\begin{figure}[H]
  \centering
  \includegraphics[width = \columnwidth]{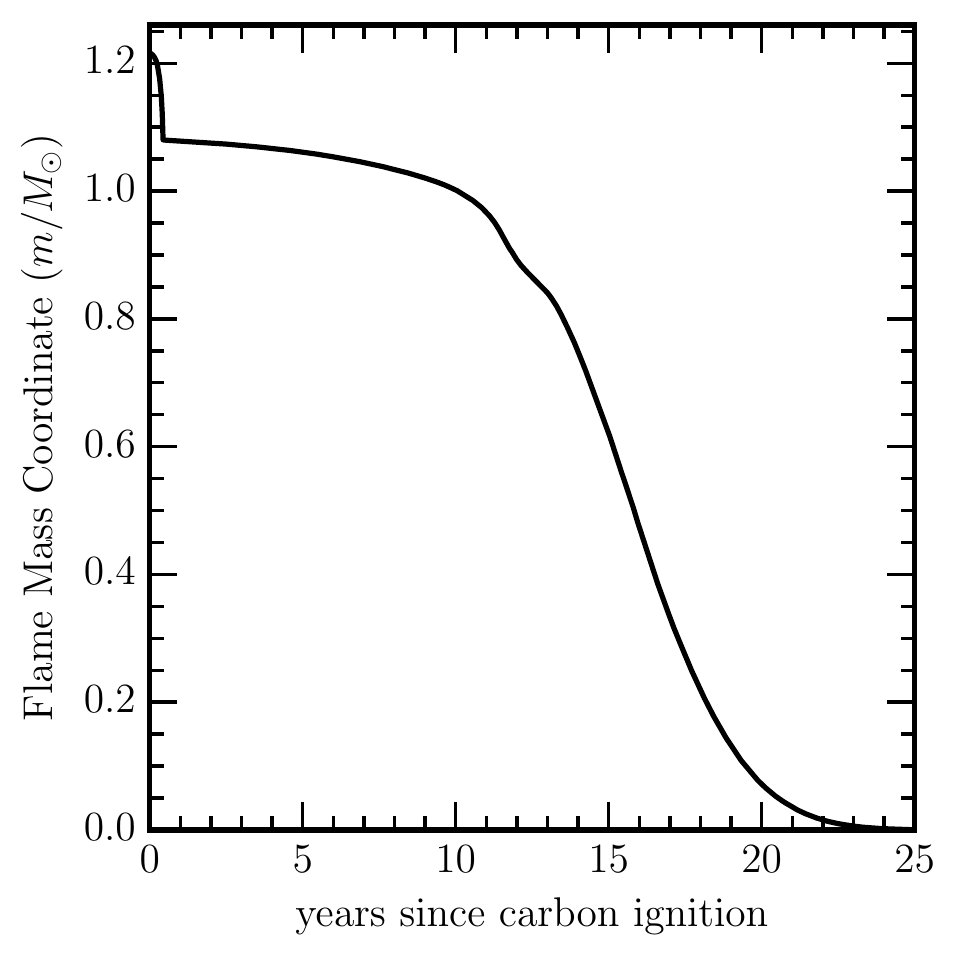}
  \caption{\footnotesize The mass coordinate of the inwardly-propagating carbon flame in a mixed hybrid C/O/Ne WD model.
  The flame speed is faster in the hot C/O ashes from steady helium burning, then it slows as it enters the mixed core which is colder and has a lower mass fraction of carbon, but then speeds up as it reaches higher temperatures and densities below the mass coordinate of $0.8 M_\odot$.}
  \label{fig:16}
\end{figure}

The complications above make it prohibitive to fully follow the flame calculations all the way to the center of C/O WDs.\footnote{An additional complication is the suggestion for normal carbon flames that
convective boundary mixing would quench the carbon flame and prevent the full conversion of any C/O  WD to O/Ne \citep{Denissenkov2013}, resulting in a WD with a cold C/O core and a hot O/Ne mantle \citep{Doherty2010, Denissenkov2013, Chen2014, Wang2014, Denissenkov2015, Farmer2015}. 
A recent multi-D study of convectively-bounded carbon flames suggests that the buoyancy barrier across the flame is too great to permit sufficient mixing to quench the flame \citep{Lecoanet2016}.
If future work shows that these flames do quench before reaching the center, then these models would not evolve towards AIC.
In \citet{Brooks2016a}, we showed that such a `hybrid' WD is in fact unstable to convection as it cools.  
However, in the accretion scenario considered in this paper, the relatively short timescale to grow to $M_{\rm Ch}$ ($\lesssim$ Myr) means that the WD would not be fully mixed at the time it nears $M_{\rm Ch}$.
A partially mixed model may experience core carbon ignition yielding a peculiar thermonuclear explosion.}
Therefore, we show here an illustrative example that contains much of the same physics of carbon flames.
We start with a $1.1M_\odot$ WD that, due to convective boundary mixing, had a $0.4M_\odot$ cold C/O core beneath a hot O/Ne mantle. 
As in \citet{Brooks2016a}, we then allowed for complete mixing over 10 Myr and then placed this WD in a 3 hr orbital period binary with a $1.5 M_\odot$ helium star.
When the WD reaches $1.25 M_\odot$ via accretion and steady helium burning, a carbon flame in the hot C/O ashes ignites and propagates into the star, as shown in Figure \ref{fig:16}. 
When the flame finishes burning through the hot C/O ashes and meets the colder, mixed hybrid core ($X_{^{12}C}\approx0.15$) the flame speed slows by almost two orders of magnitude, but continues propagating all the way to the core, converting the mixed hybrid C/O/Ne into a hot O/Ne WD in about 25 years.
This is much faster than the ${\approx}2\times10^4$ years for carbon flames in SAGB stars \citep{Timmes1994, Farmer2015}, due to the higher density of our WD interiors. 
However, even this calculation still took approximately 15,000 CPU-hours (corresponding to a wall time of months).
Since there is negligible mass loss from this event, the system will continue with mass transfer onto the newly created O/Ne WD, experience additional carbon flashes as described in \S \ref{sec:theflashes}, and eventually approach $M_{\rm Ch}$ and reach AIC conditions.
It appears likely that this fate will be generic for a range of C/O WDs accreting from sufficiently massive He star donors.

\section{Structure at AIC}\label{sec:structure}

\begin{figure}[H]
  \centering
  \includegraphics[width = \columnwidth]{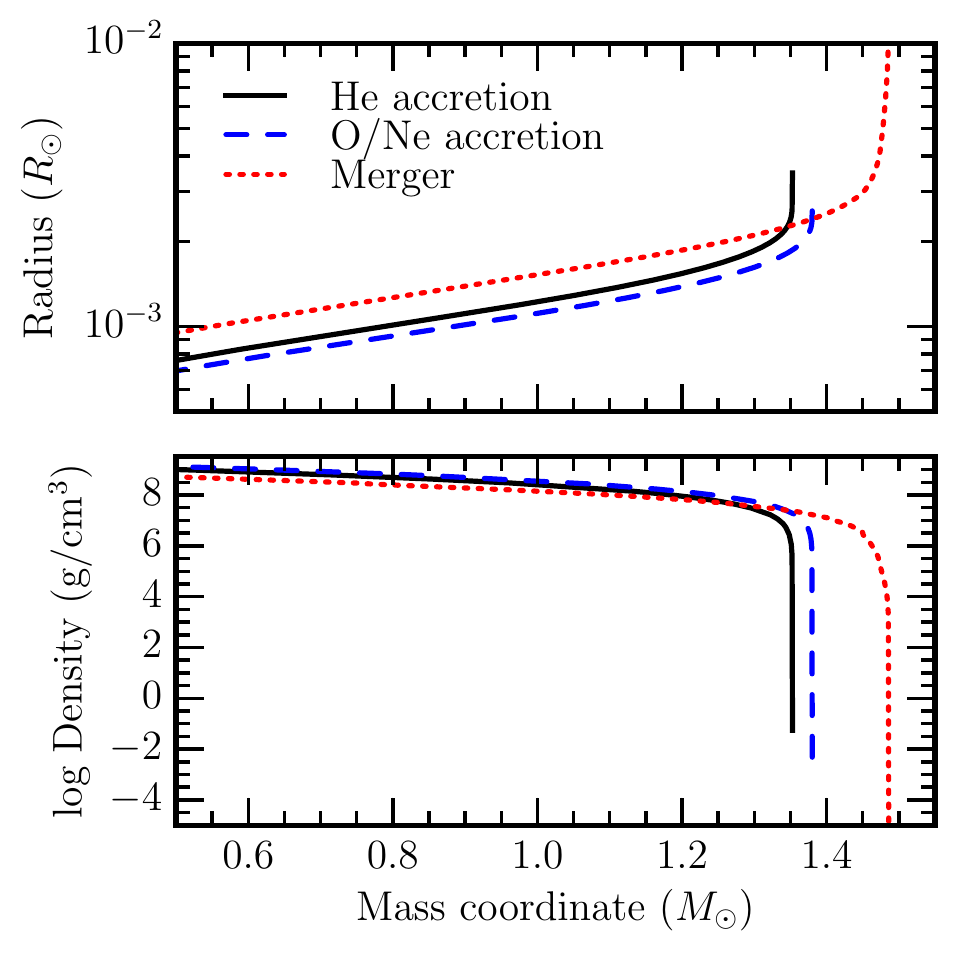}
  \caption{\footnotesize The solid black profile is from the model that starts at $1.1 M_\odot$ and has $\log\rho_{c}=9.6$.
  The blue dashed profile is from \cite{Schwab2015a} that was created as a $1.325 M_\odot$ O/Ne WD that accreted the same O/Ne mixture onto the surface until $\log\rho_{c}=9.6$.
  The red dotted profile is from \cite{Schwab2016} which is the result of a merger between two C/O WDs of masses $0.6 M_\odot$ and $0.9 M_\odot$.
  This model never reached a central density of $\log\rho_{c}=9.6$, and is instead taken after the Ne-flame has reached the center and off-center Si burning has started.
  }
  \label{fig:2}
\end{figure}

The carbon burning episodes that lead up to AIC in He star channels are important for understanding the structure of the outer layers of the WD just before AIC begins.
The observational signatures of a collapse are strongly dependent on the amount of mass ejected, and both \cite{Darbha2010} and \cite{Woosley1992} use fiducial ejected masses of $10^{-2} M_\odot$.
Figure \ref{fig:2} shows the radius versus mass and density profiles of three AIC progenitors: the model that starts as a $1.2 M_\odot$ O/Ne WD when $\log\rho_{c}=9.6$ (solid black), a model from \cite{Schwab2015a} that was created as a $1.325 M_\odot$ O/Ne WD that artificially accreted the same O/Ne mixture onto the surface at a constant rate of $10^{-6} M_\odot$/yr until $\log\rho_{c}=9.6$ (blue dashed), and the evolved remnant of the merger of two C/O WDs of masses $0.6 M_\odot$ and $0.9 M_\odot$ that is also approaching AIC (red dotted); this profile is taken from late in the evolution after carbon burning and neon-oxygen burning flames have reached the center and off-center Si ignition has occured \citep[see][for details of this model and its evolution]{Schwab2016}. 
The density profiles are similar in the core implying that the resulting accretion onto the newly formed NS and the initiation of explosion are likely to be similar.   
The merger model, however, has a larger envelope of $\approx0.2 M_\odot$ that may be ejected in the explosion, leading to a more visible electromagnetic source.   
Detailed calculations of explosions and lightcurves for both the WD merger and binary evolution AIC models would be valuable in providing insight into whether/how the observational outcome of AIC depends on the progenitor.

\begin{figure}[H]
  \centering
  \includegraphics[width = \columnwidth]{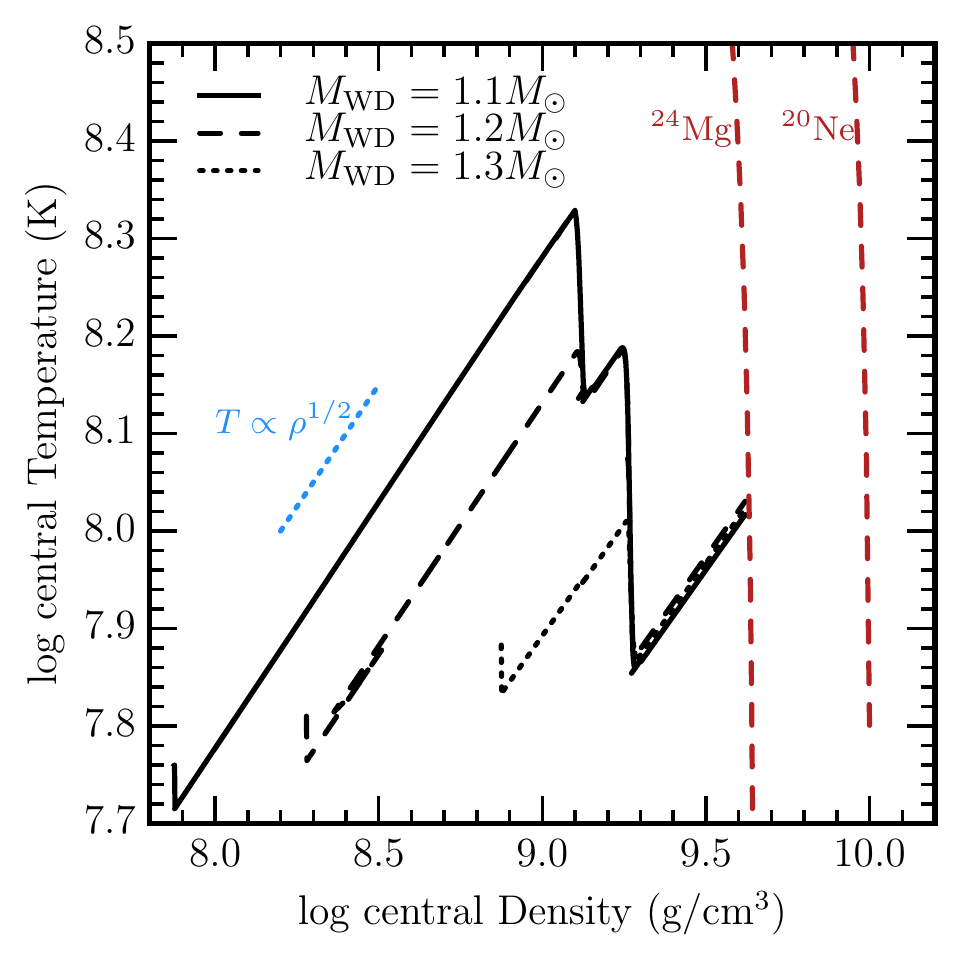}
  \caption{\footnotesize The black solid, dashed, and dotted lines show the evolution of the central density and temperature of the initially $1.1, 1.2,$ and $1.3 M_\odot$ WD models.
  The dark red dashed lines are the ignition curves for $^{24}$Mg and $^{20}$Ne electron captures.
  All models have similar temperatures at the onset of $^{24}$Mg electron captures (see text for discussion of the temperature evolution).
  }
  \label{fig:1}
\end{figure}

We find that for all of the models in this paper, the outer $10^{-2} M_\odot$ is extremely similar, due to being built up by steady helium burning on very similar core masses.
The central region should also be almost identical (shown in Figure \ref{fig:1}).  
Initially, the cores of these WDs are being adiabatically compressed on a timescale faster than the neutrino cooling timescale, leading to evolution where $T\propto\rho^{1/2}$ (as shown by the light blue dotted line), but along different adiabats.  
However, at higher densities, neutrino cooling from the ($^{25}$Mg, $^{25}$Na) and ($^{23}$Na,$^{23}$Ne) Urca pairs \citep {Paczynski1973} causes the models to evolve to the same temperature before electron captures on $^{24}$Mg begin.
The similarity of our different models, independent of initial WD mass, implies that any resulting explosion from the AIC is likely to appear very similar observationally.

\section{Conclusions}\label{sec:concl}

We have presented the first full binary simulations of He star + WD systems that lead to AIC.
We followed the standard scenario of an O/Ne WD that grows to $M_{\rm Ch}$, as well as discussing the scenario suggested by \citet{Brooks2016} in which a C/O WD accreting He experiences a carbon shell ignition, quiescently transforms into an O/Ne WD, and subsequently grows to $M_{\rm Ch}$.
Both scenarios involve helium accretion onto the WD at rates that allow steady helium burning into hot C/O ashes, a shell ignition carbon flame that propagates inwards until all the star's carbon is exhausted, followed by a series of smaller carbon shell flashes as the WD core grows in mass. 
Computational limitations prevented us from propagating the carbon flames fully through our C/O WD accretors, though the work of \cite{Timmes1994} suggests that they will reach the center.  
Future work could more directly model this conversion process.

The carbon burning flames and flashes are non-explosive and radiate a significant portion of their energy in neutrinos.
The flashes cause the radius and luminosity of the WD to increase by 25-100\%, but are separated by at least hundreds of years, so they may be difficult to observe.

We evolve our models until they reach central densities at which electron captures will begin to occur, but do not follow them to collapse.\footnote{Recently, using multi-dimensional simulations of the oxygen deflagration, \citet{Jones2016} suggested that collapse to a NS may not be assured.}
The capability to evolve realistic models up to the onset of the hydrodynamic collapse of the WD is still under active development in \texttt{MESA} and is therefore left for a future study.  The current electron capture physics employed in \texttt{MESA} is covered in \cite{Schwab2015a} for idealized models (simple compositions, steady accretion rate, no surface burning).

As shown in Figures \ref{fig:15} and \ref{fig:1}, as long as the donor has enough helium for the WD to grow to $M_{\rm Ch}$ (taking into account wind losses), the conditions of the models at AIC are very similar over a wide range of initial orbital periods and WD masses.

There have been no direct observations of AICs.
They are predicted to be faint and very short lived \citep{Woosley1992, Dessart2006} and so they may be very difficult to observe. 
Unusual NSs provide the best evidence for AIC to date, although it is rather indirect.
For example, we may see evidence for these events in the form of recycled MSPs \citep{Tauris2013}.
In addition, if NS formation via AIC does not cause a kick, then they could explain the large numbers of NSs in globular clusters, and the subsequent spin-up from accretion could make them look young \citep{Boyles2011, Antoniadis2016}.
In our calculation, the model with the $1.1 M_\odot$ WD undergoes AIC when the donor only has $0.5 M_\odot$ of helium left, most of which will be ejected due to the fact that the mass loss rate from the donor is a few orders of magnitude higher than the Eddington limit for NSs. 
This leaves little mass for spin-up accretion post-AIC, meaning systems like this may be the progenitors to high B-field, slow spinning NSs that are seen in globular clusters \citep{Tauris2013}.
\citet{Bailyn1990} show that their population study suggest that NSs born via AIC may resolve discrepancies in calculated birthrates.

\cite{Yungelson1998} and \cite{Kwiatkowski2015} give AIC rates from He star systems in spiral galaxies as a few $\times10^{-5}$ per year, which should be significantly increased if our newly revealed C/O to O/Ne to AIC channel proves effective.
Furthermore, \cite{Kwiatkowski2015} show that the Helium star channel rate is comparable to the channels with Hertzsprung gap star, Red Giant star, AGB star, and WD donors.
The inclusion of the C/O to O/Ne AIC channel may therefore significantly increase the predicted AIC rate due to the fact that lower mass C/O WDs are inherently more common than higher mass O/Ne WDs.

This research is funded in part by the Gordon and Betty Moore Foundation through Grant GBMF5076 to L.B. and E.Q..
We acknowledge stimulating workshops at Sky House where these ideas germinated. 
This work was supported by the National Science Foundation under grant PHY 11-25915. 
J.S. was supported in part by the National Science Foundation Graduate Research Fellowship Program under Grant No. DGE 11-06400 and by NSF Grant No. AST 12-05732. 
E.Q. was supported in part by a Simons Investigator award from the Simons Foundation and the David and Lucile Packard Foundation. 
Support for this work was provided by NASA through Hubble Fellowship grant \# HST-HF2-51382.001-A awarded by the Space Telescope Science Institute, which is operated by the Association of Universities for Research in Astronomy, Inc., for NASA, under contract NAS5-26555.
Most of the simulations for this work were made possible by the Triton Resource, a high-performance research computing system operated by the San Diego Supercomputer Center at UC San Diego.

\bibliographystyle{apj}
\bibliography{aic}

\end{document}